\documentclass[twocolumn]{aastex631}

\shorttitle{Dark age of Type II SNRs}
\shortauthors{Yasuda et al.}
\graphicspath{{./}{figures/}}

\begin{document}

\title{Dark Age of Type II Supernova Remnants}

\author[0000-0002-0802-6390]{Haruo Yasuda}
\affiliation{Department of Astronomy, Kyoto University, \\
Kitashirakawa, Oiwake-cho, Sakyo-ku, Kyoto 606-8502, Japan}

\author[0000-0002-2899-4241]{Shiu-Hang Lee}
\affiliation{Department of Astronomy, Kyoto University, \\
Kitashirakawa, Oiwake-cho, Sakyo-ku, Kyoto 606-8502, Japan}
\affiliation{Kavli Institute for the Physics and Mathematics of the Universe (WPI), The University of Tokyo, Kashiwa 277-8583, Japan}

\author[0000-0003-2611-7269]{Keiichi Maeda}
\affiliation{Department of Astronomy, Kyoto University, \\
Kitashirakawa, Oiwake-cho, Sakyo-ku, Kyoto 606-8502, Japan}

\correspondingauthor{Haruo Yasuda}
\email{yasuda@kusastro.kyoto-u.ac.jp}



\begin{abstract}

Supernova remnants (SNRs) are important objects in terms of their connections with supernova (SN) explosion mechanism(s), progenitor stars, and cosmic-ray acceleration. Non-thermal emission from SNRs is an effective probe of the structure of their surrounding circumstellar media (CSM), which can in turn shed lights on mechanism and history of the elusive mass-loss of massive stars. In this work, we calculate the time evolution of broadband non-thermal emission from SNRs originating from Type II SNe embedded in a CSM environment linked to the mass loss history of the progenitor. Our results predict that Type II SNRs experience a prolonged period of weak radio and $\gamma$-ray emission if they run into a spatially extended bubble of low density and high temperature created by the stellar wind during main sequence. For a typical red supergiant progenitor evolved within an average interstellar medium (ISM), this ``dark age" corresponds to a range of SNR ages spanning from $\sim 1000$ to $5000$~yrs old. This result suggests that a majority of Type II SNRs are too faint to be detected, which may help explain why the number of known Galactic SNRs is significantly less than what we expect from the SN rate in our Galaxy. 

\end{abstract}



\section{Introduction}\label{sec:Intro}
Supernovae (SNe) are one of the most energetic phenomena in the Universe in which stars explode and release a tremendous amount of energy at the final stage of stellar evolution. Type II SNe are known to be coming from the death of massive stars in their final evolutionary stage such as red supergiants (RSG) \citep{2015PASA...32...16S}. Electromagnetic radiation from SNe provides information about their progenitors and surrounding environments, which are crucial in understanding stellar evolution and mass loss history of massive stars \citep{1997ARA&A..35..309F}. However, SN observations are usually limited to a timescale of an order of weeks to years, which means that we can only extract the mass loss history shortly before explosion. On the other hand, observations of their supernova remnants (SNRs) interacting with their CSM environments are an effective supplementary tool for probing mass loss at earlier phases well before core collapse.

Young and dynamically active SNRs are usually observable in multi-wavelength from radio to TeV-$\gamma$ rays, indicating that SNRs are in-situ acceleration sites of relativistic particles, which are widely believed to be closely linked to the origin of Galactic cosmic rays (CRs) accelerated at the SNR shock fronts through the diffusive shock acceleration (DSA) mechanism \citep{1949PhRv...75.1169F,1978MNRAS.182..147B,1978ApJ...221L..29B}. The non-thermal emissions are mostly produced by the interactions between the accelerated CRs and the surrounding interstellar medium (ISM) and circumstellar medium (CSM). They therefore hold the key to understanding the ambient environments in which SNe explode. \cite{2019ApJ...876...27Y} (hereafter YL19) calculated the evolution of young SNRs and the accompanying non-thermal emissions in various environments until 5000 yr, and they found that the spectral energy distribution (SED) of the broadband emission varies with time in a way strongly correlated with the density and spatial structure of the surrounding ISM/CSM gas and magnetic field. However, they used very simplified models for the environments by assuming simple power-law distributions extended to infinity for the CSM density for example, without considering the mass loss history and stellar evolution of the progenitor stars. A systematic calculation linking the progenitors, SNe and SNRs, especially with the mass loss history taken into account, is therefore on high demand for facilitating the usage of SNR observations for diagnosing SN types, mass loss mechanism and progenitor natures.

In this study, we first prepare realistic CSM models using one-dimensional hydrodynamic simulations considering the mass-loss history of a Type II SN progenitor. Using another set of hydrodynamical simulations coupled with efficient particle acceleration, we then compute the time evolution of SNR dynamics and non-thermal emissions in such CSM environments up until an age of $10^4$ yrs. In section.~\ref{sec:Method}, we introduce our numerical method for the hydrodynamics and particle acceleration for SNR evolution, and for the generation of reasonable CSM models based on SN observations. Section.~\ref{sec:Results} shows our results on the non-thermal emissions from SNRs assuming different progenitor masses and stellar wind properties, and their comparisons to the currently available observation data. Discussions and conclusion are summarized in Section.~\ref{sec:Discussion} and Section.~\ref{sec:Conclusion}  

\section{Method}\label{sec:Method}

\subsection{Hydrodynamics} \label{physics}

The hydrodynamics code used in this work is in a large part identical to the {\it CR-Hydro} code developed in YL19 except for a few differences which we will overview in the following. The hydrodynamic calculations are based on the {\it VH-1} code \citep[e.g.,][]{2001ApJ...560..244B} which solves multi-dimensional Lagrangian hydrodynamic equations. As introduced in YL19, we modified the code to include feedback from CR acceleration, and assumed a spherical symmetry for simplicity; 
\begin{eqnarray}
&&\frac{\partial r}{\partial m} + \frac{1}{4\pi r^2\rho} = 0\\
&&\frac{\partial u}{\partial t} + \frac{\partial P_\mathrm{tot}}{\partial m} = 0\\
&&\frac{\partial e}{\partial t} + \frac{\partial }{\partial m}(P_\mathrm{tot}u) = -n^2\Lambda_\mathrm{cool}\\
&&e=\frac{1}{2}u^2+\frac{P_\mathrm{tot}}{(\gamma_\mathrm{eff}-1)\rho},
\end{eqnarray}
where $\rho$, $n$, $m$, $u$ and $e$ are the gas mass density, number density, mass coordinate, fluid velocity and internal energy density, respectively. 
We treat the gas and accelerated CRs in an one-fluid description by employing an effective gamma $\gamma_\mathrm{eff}$ for the equation-of-state \citep[e.g.,][]{1983ApJ...272..765C,2001ApJ...560..244B}, and a total pressure defined as $P_\mathrm{tot}=P_\mathrm{g}+P_\mathrm{CR}+P_\mathrm{B}$, where $P_\mathrm{g}$, $P_\mathrm{CR}$ and $P_\mathrm{B}$ are gas pressure, CR pressure and magnetic pressure, respectively. Since {\it VH-1} is not a magnetohydrodynamics (MHD) code, we provide an additional treatment for the time evolution of the post-shock magnetic field strength. Ignoring effects such as amplification by MHD turbulence, the magnetic field strength follows the conservation of magnetic flux $B\propto r^{-2}$ along with the advection of the downstream gas. Combined with the mass conservation $\rho\propto r^{-2}$, we can obtain $B\propto\rho$.  As in YL19, the magnetic field also receives an amplification by CR-streaming instability in the shock precursor which is calculated self-consistently with the particle acceleration. The temperatures of protons $T_p$ and electrons $T_e$ are equilibrated by their post-shock Coulomb collisions. To allow for the calculation of late-phase SNR evolution, especially in a high-density medium which YL19 did not consider, we implement optically thin radiative cooling as well in this work using an exact integration scheme \citep{2009ApJS..181..391T}. A non-equilibrium ionization cooling curve from \cite{1993ApJS...88..253S} is used for the cooling function $\Lambda_\mathrm{cool}$. 

\subsection{Cosmic-ray spectrum}
The phase-space distribution function of the accelerated protons, $f_p(x,p)$, can be obtained by solving the following diffusion-convection equation written in the shock rest frame \citep[e.g.,][]{2010APh....33..307C,2010MNRAS.407.1773C,LEN2012} assuming a steady-state and isotropic distribution in momentum space,
\begin{eqnarray}\label{eq:DSA}
&&[u(x)-v_A(x)]\frac{\partial f_p(x,p)}{\partial x}-\frac{\partial}{\partial x}\left[D(x,p)\frac{\partial f_p(x,p)}{\partial x}\right] \nonumber\\
&&=\frac{p}{3}\frac{d[u(x)-v_A(x)]}{dx}\frac{\partial f_p(x,p)}{\partial p}+Q_p(x,p) ,
\end{eqnarray}
where $D(x,p)$, $v_A(x)$ and $Q_p(x,p)$ is the spatial diffusion coefficient, Alfv\'en speed and proton injection rate at position $x$ in the shock rest frame. We assume a Bohm diffusion, such that $D(x,p)=pc^2/3eB(x)$, where $B(x)$ is the local magnetic field strength at position $x$. We adopt the so-called `thermal-leakage' injection model \citep{2004APh....21...45B,2005MNRAS.361..907B} for the DSA injection rate $Q_p(x,p)$ such that 
\begin{equation}
Q_p(x,p) = \eta\frac{n_{1}u_1}{4\pi p_{\mathrm{inj}}^2}\delta(x)\delta(p-p_{\mathrm{inj}}),
\end{equation}
where $n_{1}$ is the number density of proton at immediately upstream of the shock and $p_{\mathrm{inj}}$ is the CR injection momentum, which is defined as $p_{\mathrm{inj}} = \chi_\mathrm{inj}\sqrt{2m_pk_\mathrm{b}T_p}$, where $m_p$, $k_\mathrm{B}$ and $T_p$ are the proton mass, Boltzmann constant and temperature respectively. $\chi_\mathrm{inj}$ and $\eta$ are free parameters in this work, which control the fraction of thermal particles injected into the DSA process as described in YL19.  

Here, we solve eq.~(\ref{eq:DSA}) at the shock position $x=0$ so that the distribution function can be written in an implicit form with an exponential cutoff \citep{2004APh....21...45B,2005MNRAS.361..907B}; 
\begin{eqnarray}
f_{p}(x=0,p) &=& \frac{\eta n_{0}}{4\pi p_{\mathrm{inj}}^3}\frac{3S_\mathrm{tot}}{S_\mathrm{tot}U(p)-1} \nonumber\\
&&\times\exp\left(-\int_{p_{\mathrm{inj}}}^p \frac{dp'}{p'}\frac{3S_\mathrm{tot}U(p')}{S_\mathrm{tot}U(p')-1}\right) \nonumber\\
&&\times \exp\left[-\left(\frac{p}{p_\mathrm{max,p}}\right)^{\alpha_\mathrm{cut}}\right],
\end{eqnarray}
where $S_\mathrm{tot}$ and $U(p)$ are the effective compression ratio and normalized fluid velocity, respectively. The explicit expressions of these quantities are easily obtained by referring to \cite{2010APh....33..307C} and \cite{LEN2012}. $\alpha_\mathrm{cut}$ is introduced because of a poor understanding of the escape process of CRs, which is directly related to the CR spectral shape beyond the maximum momentum $p_\mathrm{max,p}$.

For the electron spectrum, we use a parametric treatment where the electron distribution function is given as $f_e(x,p) = K_{ep}f_p(x,p)\exp[-(p/p_\mathrm{max,e})^{\alpha_\mathrm{cut}}]$. $K_{ep}$ typically takes a value between $10^{-3}$ and $10^{-2}$ based on constraints from SNR observations. The determination of the maximum momenta of each particle species is the same as in YL19.

The particles accelerated at the shock is assumed to be co-moving with the gas flow and suffer from energy loss through non-thermal radiations and adiabatic loss. For the non-thermal radiation mechanisms, we consider synchrotron radiation, inverse Compton scattering (IC), bremsstrahlung from the accelerated electrons, and pion productions by proton-proton interaction ($\pi^0$ decay) by the accelerated protons.

\subsection{Circumstellar medium and SN ejecta}\label{Method:CSM}
In this study, we first prepare models for the circumstellar medium (CSM) of a Type II SNR by accounting for stellar evolution and mass loss histories of the SN progenitor. The CSM models are generated by performing hydrodynamic simulations in which stellar winds run into a uniform ISM region. The results are used as the initial conditions for the subsequent calculation for the evolution of the SNR.   

The progenitor of a Type II SN is believed to be massive OB stars with zero-age main sequence (ZAMS) mass $\ge10M_\odot$. This type of stars evolves to red supergiants (RSG) after their main sequence (MS) phase, and explodes via core collapse of their iron cores. Although the mass loss mechanism is not well understood and is still under discussion, it is thought that the star loses its mass from its envelope mainly in the form of stellar wind. The wind blown in MS phase is thin and fast from the compact OB stars, and the total amount of mass lost in the MS phase is relatively small. On the contrary, the star loses most of its mass in the RSG phase through a denser and slower wind. The typical values for the mass loss rate $\dot{M}_\mathrm{w}$, wind velocity $V_\mathrm{w}$ and time duration $\tau_\mathrm{phase}$ in each phase are, $\dot{M}_\mathrm{w}\sim 10^{-8}-10^{-7}\ M_\odot/\mathrm{yr}$, $V_\mathrm{w}\sim1-3\times10^3\ \mathrm{km/s}$ and $\tau_\mathrm{phase}\sim 10^6-10^7\ \mathrm{yr}$ for the MS phase, and $\dot{M}_\mathrm{w}\sim 10^{-6}-10^{-5}\ M_\odot/\mathrm{yr}$, $V_\mathrm{w}\sim10-20\ \mathrm{km/s}$ and $\tau_\mathrm{phase}\sim 10^5-10^6\ \mathrm{yr}$ for the RSG phase. The relation between the ZAMS mass and pre-SN mass of the progenitor has been investigated \citep[e.g.,][]{2009ApJ...703.2205K,2014ApJ...783...10S,2015ApJ...810...34W,2016ApJ...821...38S}, so the mass lost through the MS and RSG winds, and the ejecta mass $M_\mathrm{ej}$ can be determined if the ZAMS mass is fixed. In this study, we consider two cases for the ZAMS mass, i.e., a $12M_\odot$ (model A) and $18M_\odot$ (model B) progenitor star. We also use a time-independent, constant mass loss rate and wind velocity during each phase for simplicity. The exact values used in the models are summarized in Table.~\ref{table:wind}.

When these progenitors explode, the stellar debris propagates outward as a SN ejecta, but some of it falls back onto the stellar core which forms a neutron star. The ejecta mass is calculated as $M_\mathrm{ej} = M_\mathrm{ZAMS} - \sum (\dot{M}_\mathrm{w}\tau_\mathrm{phase}) - M_\mathrm{rm}$, where $M_\mathrm{rm}$ is the stellar remnant mass after explosion. In the ZAMS mass range we consider in this work, $M_\mathrm{rm}$ is typically $1.4\sim1.7\ M_\odot$ \citep[]{2007PhR...442..269W,2016ApJ...821...38S,2020ApJ...896...56W}.  $M_\mathrm{rm}=1.5\ M_\odot$ is adopted in all models here.
For the SN ejecta structure, we assume a power-law envelope model in \citet[]{1999ApJS..120..299T} for all of our models;
\begin{eqnarray}\label{eq:ejecta}
\rho(r)=&&\rho_\mathrm{c}\ \ \ \ \ \ \ \ \ \ \ \ \ \ (r\le r_\mathrm{c}) \nonumber\\
        &&\rho_\mathrm{c}(r/r_\mathrm{ej})^{-n_\mathrm{SN}}\ (r_\mathrm{c}\le r \le r_\mathrm{ej}),
\end{eqnarray}
where $\rho_\mathrm{c}$, $r_\mathrm{c}$ and $r_\mathrm{ej}$ are the core density, core radius and ejecta size, respectively. These values are uniquely determined by mass and energy conservation. The related parameters are, therefore, the ejecta mass $M_\mathrm{ej}$, the kinetic energy of the explosion $E_\mathrm{SN}$, and the power-law index of the envelope $n_\mathrm{SN}$. We assume $E_\mathrm{SN} = 1.2\times10^{51}\ \mathrm{erg}$ and $n_\mathrm{SN} = 7$. The ejecta masses depend on the ZAMS masses in each model, and are summarized in Table.~\ref{table:wind}.

\begin{figure}[ht!]
\plotone{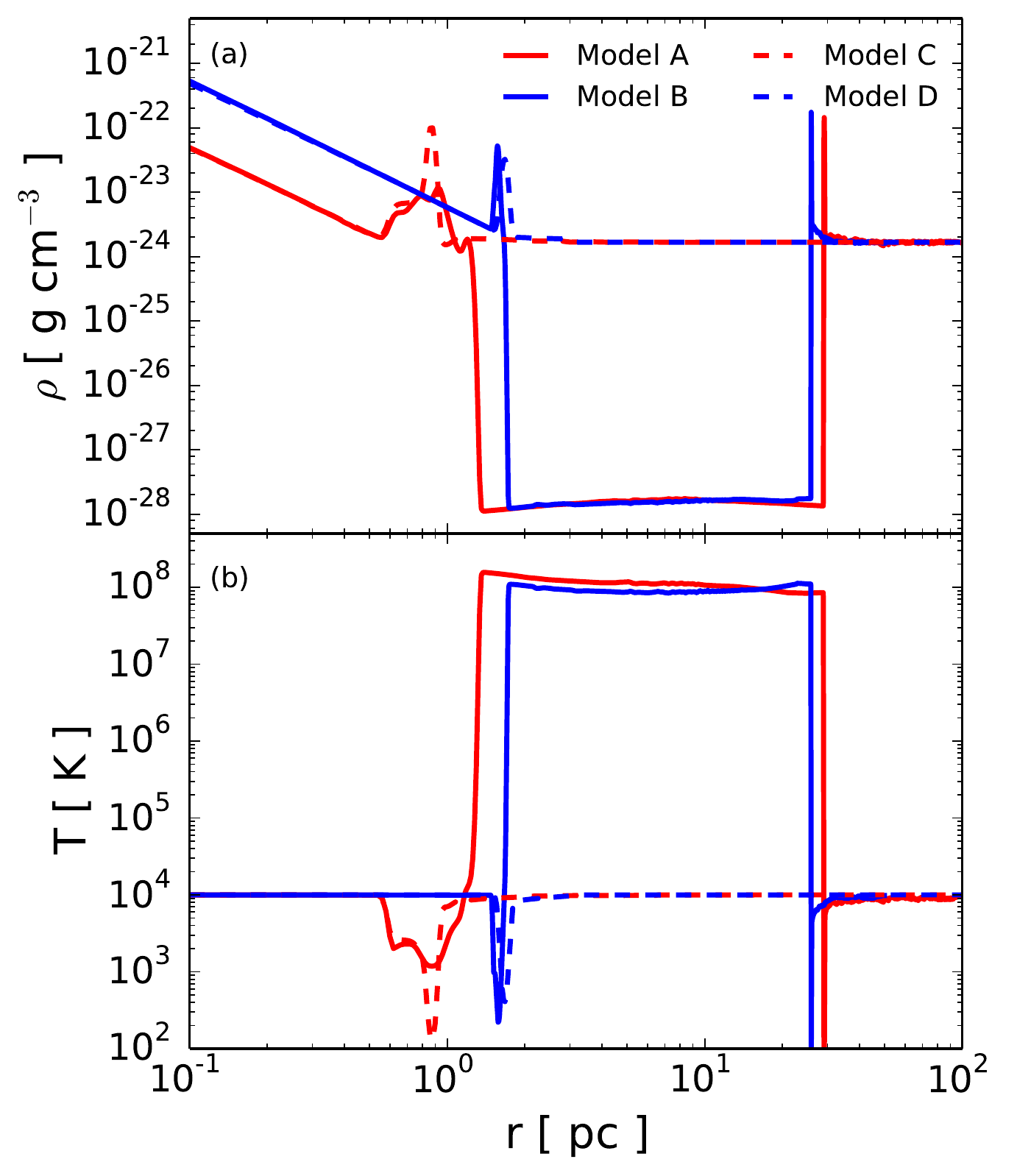}
\caption{CSM models for a Type II SNR. The upper panel shows the gas density as a function of radius, and the lower panel shows the average gas temperature. The red (blue) solid line corresponds to the low (high) progenitor mass case. The dashed lines show the results from models in which the MS bubble does not exist for comparison. \label{fig:wind}}
\end{figure}

\begin{deluxetable*}{ccccccccc}
\tablecaption{Model parameters \label{table:wind}}
\tablewidth{0pt}
\tablehead{
\colhead{Model} & \colhead{$M_\mathrm{ZAMS}$} & \colhead{wind phases} & \colhead{$\dot{M}$} & \colhead{$V_\mathrm{w}$} & \colhead{$M_\mathrm{w}$} & \colhead{$\tau_\mathrm{phase}$} & \colhead{$M_\mathrm{ej}$}\\
\colhead{} & \colhead{[$M_\odot$]} & \colhead{} & \colhead{[$M_\odot/\mathrm{yr}$]} & \colhead{[$\mathrm{km/s}$]} & \colhead{[$M_\odot$]} & \colhead{[$\mathrm{yr}$]} & \colhead{[$M_\odot$]}}
\startdata
A & 12 & MS    & $5.0\times10^{-8}$ & 2000   & 0.5 & $10^7$          &     \\
  &    & RSG   & $1.0\times10^{-6}$ & 10     & 0.5 & $5.0\times10^5$ & 9.5 \\
\hline
B & 18 & MS    & $6.0\times10^{-8}$ & 2000   & 0.3 & $5.0\times10^6$ &      \\
  &    & RSG   & $1.0\times10^{-5}$ & 10     & 2.7 & $2.7\times10^5$ & 13.5  \\
\hline
C & 12 & RSG   & $1.0\times10^{-6}$ & 10     & 1.0 & $10^6$          & 9.5  \\
\hline
D & 18 & RSG   & $1.0\times10^{-5}$ & 10     & 3.0 & $3.0\times10^5$ & 13.5
\enddata
\tablecomments{The wind parameters and ejecta properties for a Type II SNR. The wind temperature is set to be $T = 10^4\ \mathrm{K}$, SN explosion energy $E_\mathrm{SN}=1.2\times10^{51}\ \mathrm{erg}$, power-law index of the ejecta envelope $n_\mathrm{ej}=7$, and stellar remnant mass $M_\mathrm{rm} = 1.5M_\odot$ \citep[]{2007PhR...442..269W,2016ApJ...821...38S,2020ApJ...896...56W} in all models. We also assume $n=1.0\ \mathrm{cm}^{-3}$ and $T=10^4\ \mathrm{K}$ for the outer ISM region.}
\end{deluxetable*}

Figure.~\ref{fig:wind} shows the density and temperature structures provided by our stellar wind simulations. The upper panel (a) shows the radial density distribution of the CSM created by the stellar wind from a Type II SN progenitor. The red and blue solid lines correspond to the results of the $12\ M_\odot$ (model A) and $18\ M_\odot$ (model B) cases, respectively. The dashed lines represent the models in which mass loss in the MS phase is not considered for comparison (model C and D). The lower panel (b) shows the gas temperature as a function of radius. In the stellar wind simulations, the winds are assumed to be blown into a uniform ISM with $n_\mathrm{ISM}=1.0\ \mathrm{cm}^{-3}$ and $T=10^4\ \mathrm{K}$ in all of our models. 

From the solid lines in panel (a) and (b), we can see that the CSM structure can be divided into 5 characteristic regions from the outer to inner radius; (i) uniform ISM, (ii) MS shell, (iii) MS bubble, (iv) RSG shell, and (v) RSG wind. Because the MS wind has a low density and high velocity, and it is blown over a relatively long time period, the MS wind sweeps up the ISM and forms a dense cold shell between the ISM and the MS bubble at $r\sim30\ \mathrm{pc}$. The swept ISM mass is $M=(4\pi/3) r^3 m_\mathrm{p}n_\mathrm{ISM}\sim 2700\ M_\odot(r/30\ \mathrm{pc})^3(n_\mathrm{ISM}/1\ \mathrm{cm}^{-3})$, which is much larger than the total mass inside the MS wind $\sim 0.5\ M_\odot$. A termination shock is formed and heats the MS wind up to a high temperature. As a result, the environment is characterized by a tenuous ($n\sim10^{-4}\ \mathrm{cm}^{-3}$) and hot ($T\sim10^8\ \mathrm{K}$) plasma as a ``MS bubble". After that, the RSG wind sweeps up the thin gas inside the bubble, and a RSG wind shell is formed at the outer edge of the wind at $r\sim1\ \mathrm{pc}$. 

The differences between models A and B are mainly in the locations of the MS shell and RSG shell. They are attributed to the slight differences in the mass loss rates and time duration of the mass loss phases mainly determined by the mechanical balance between the ram pressure of the winds and the thermal pressure of the external gas. On the other hand, while model C and D do not include mass loss in the MS phase intentionally, the RSG shells locate at more-or-less the same radius as models A and B because the thermal pressures in the ISM and the MS bubble are almost the same. In overall, the major difference between models A and B (solid lines) and models C and D (dashed lines) lies in the (non-)existence of the MS bubble and MS shell. 

The results from the stellar wind simulations above are used as the initial conditions for our subsequent calculations for the evolution of the SNR. We further define the local magnetic field strength in the CSM environment as $B(r)=\sqrt{8\pi n(r)k_\mathrm{B}T(r)/\beta}$, where $\beta$ is the plasma beta $\beta \equiv P_\mathrm{g}/P_\mathrm{B}$. From observations of SNe and SNRs, $\beta$ is typically $\ge 100$ inside a wind, and $\sim 1$ in the ISM close to equi-partition. In this study, $\beta$ and the other free parameters mentioned above are obtained by fitting to the observation of SNR RX J1713.7-3946 as in YL19, i.e., $\beta \sim 825$ for the unshocked wind and wind shells, and $\beta \sim 2.17$ for the ISM region, which correspond to magnetic field strengths $B\sim0.3\ \mathrm{\mu G}$ in the wind region (at $r\sim1\ \mathrm{pc}$) and $B\sim4.0\ \mathrm{\mu G}$ in the ISM region (at $r\ge10\ \mathrm{pc}$). The other parameters such as $\chi_\mathrm{inj}$ and $\alpha_\mathrm{cut}$ are the same as in Model B in YL19.

\section{Results} \label{sec:Results}
In the SNR simulations, we compute the hydrodynamical evolution of a Type II SNR up to an age of $10^4$ yr, and the non-thermal emissions resulted from its interaction with the environment models provided by the wind simulations as described in the previous section.

\begin{figure}[ht!]
\plotone{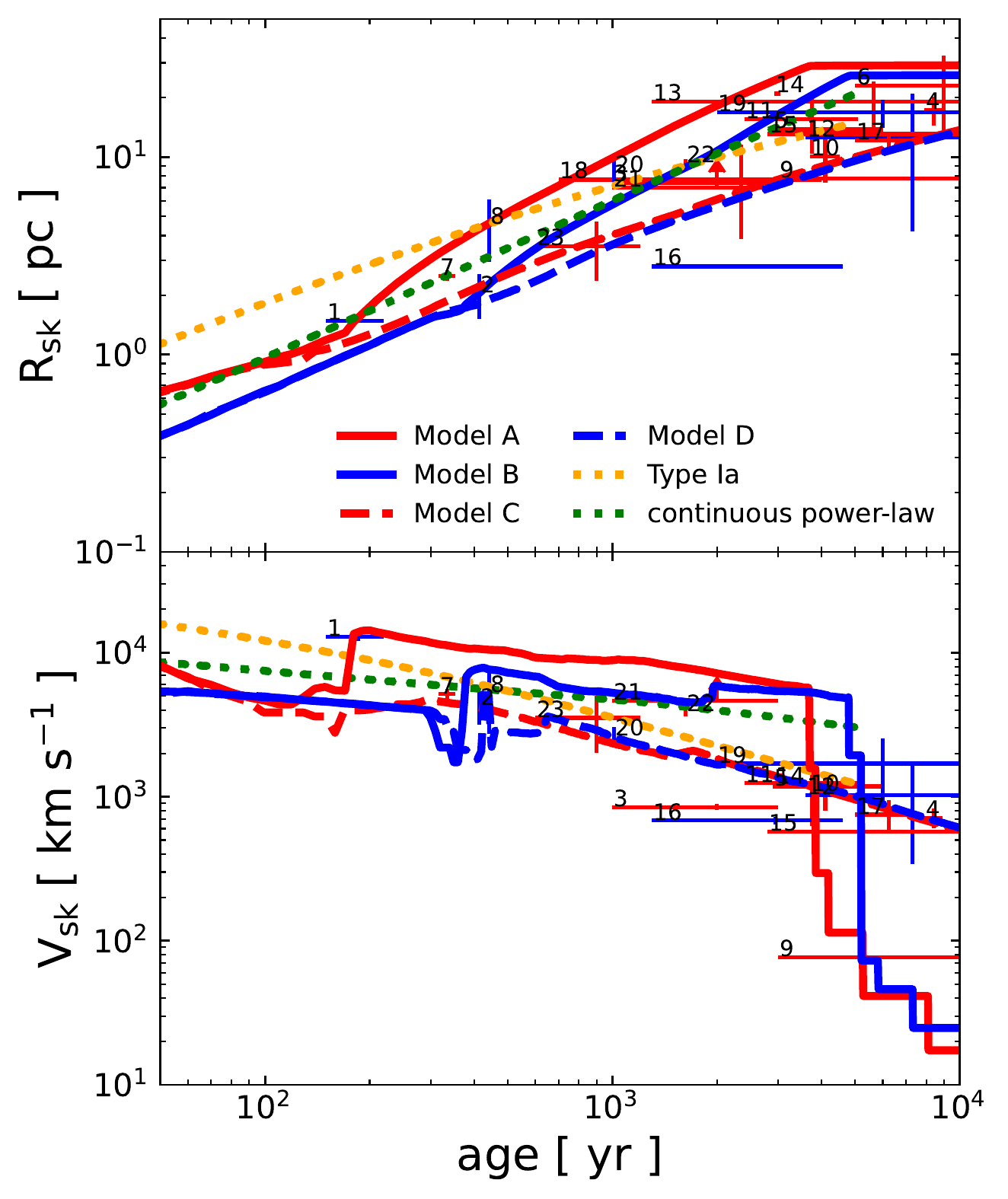}
\caption{The hydrodynamical evolution of a Type II SNR. Upper panel shows  the forward shock radius as a function of SNR age, and the lower panel shows the evolution of the shock velocity. The line formats are the same as Figure.~\ref{fig:wind}. The dotted lines are taken from Model A2 (orange) and B2 (green) in YL19 for comparison (see text). Actual observation data from $\gamma$-ray bright SNRs are overlaid, for which the references can be found in YL19. \label{fig:RV_t}}
\end{figure}

\begin{figure*}[ht!]
\gridline{\fig{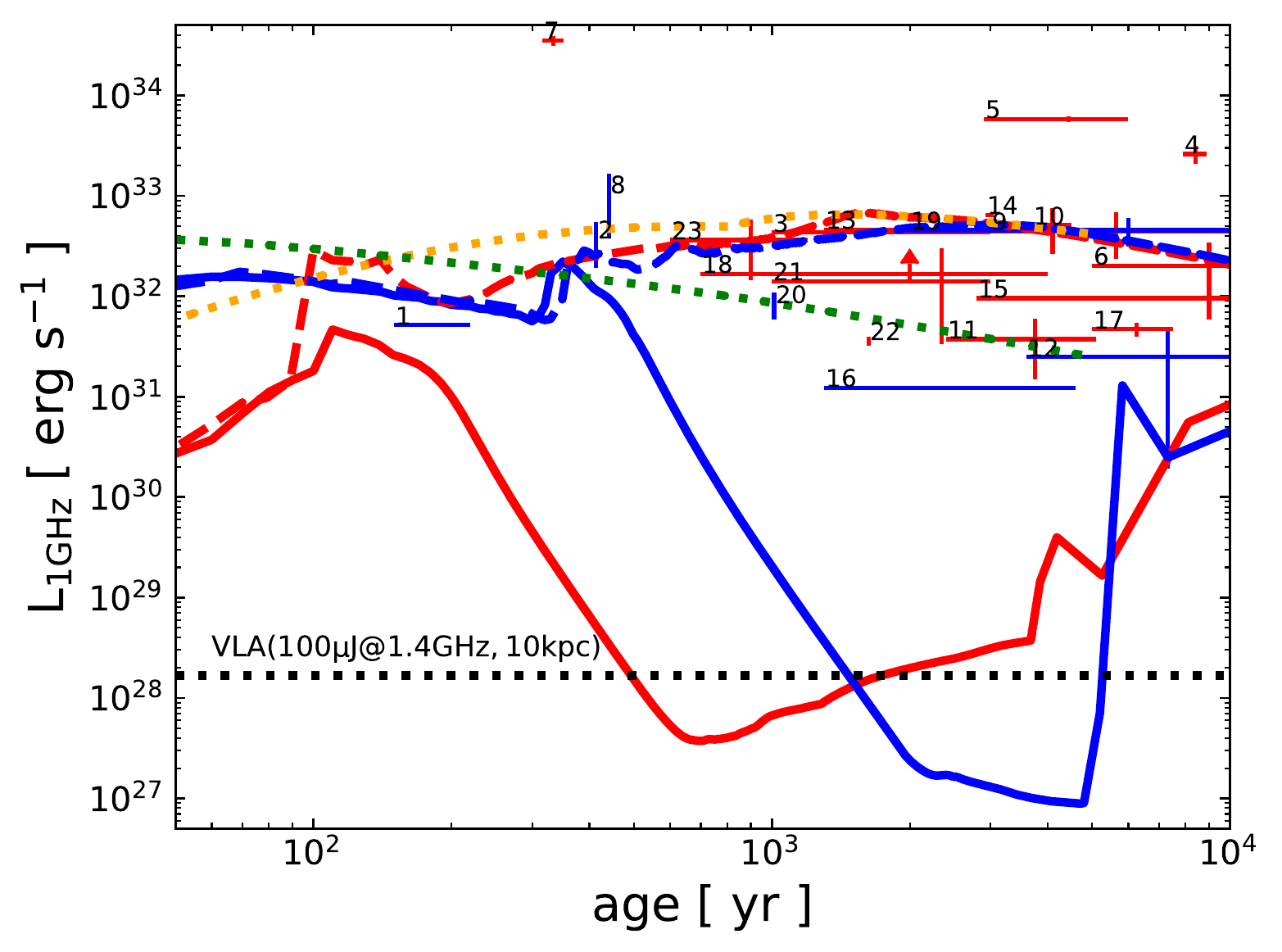}{0.33\textwidth}{(a)}
          \fig{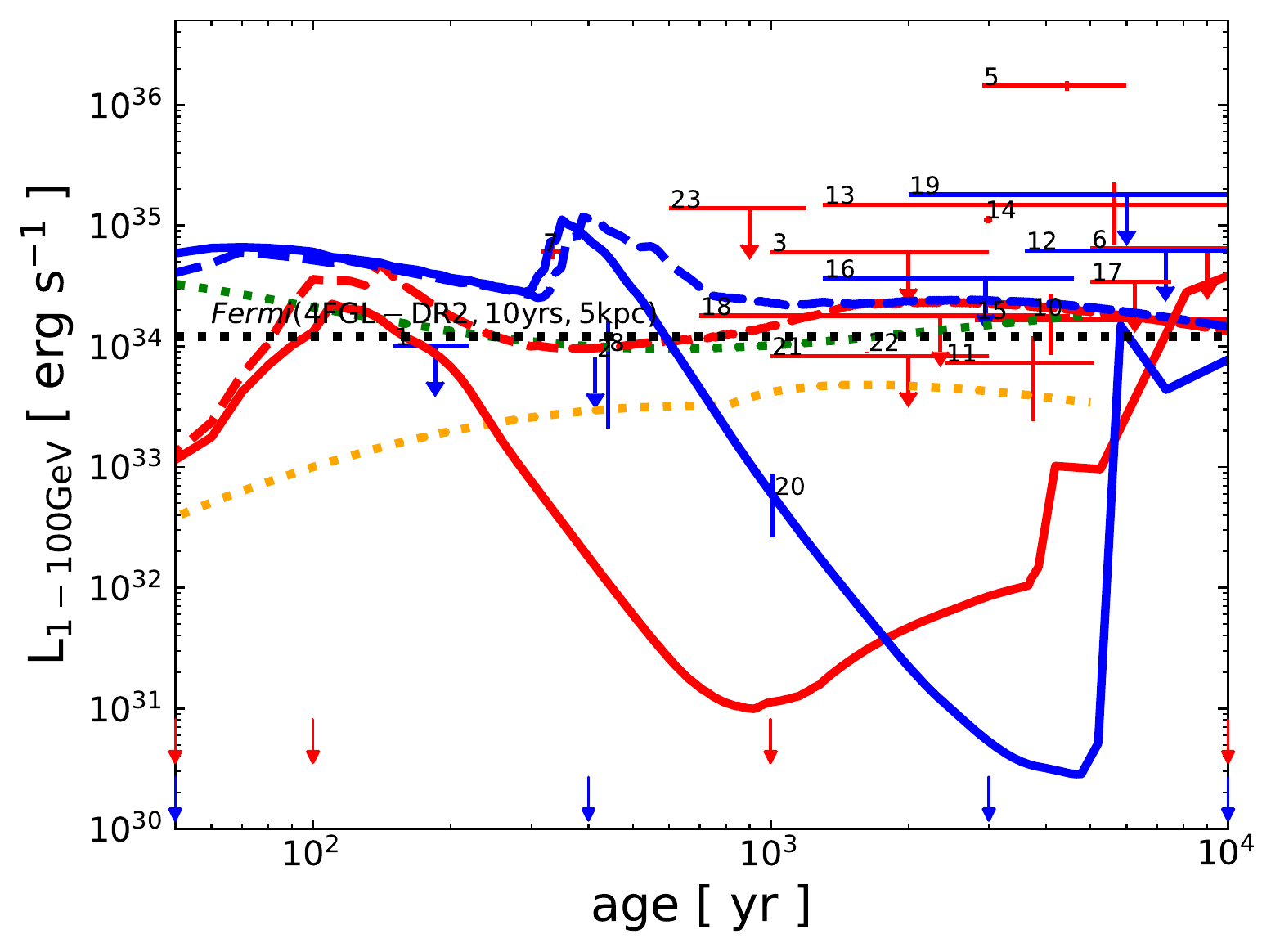}{0.33\textwidth}{(b)}
          \fig{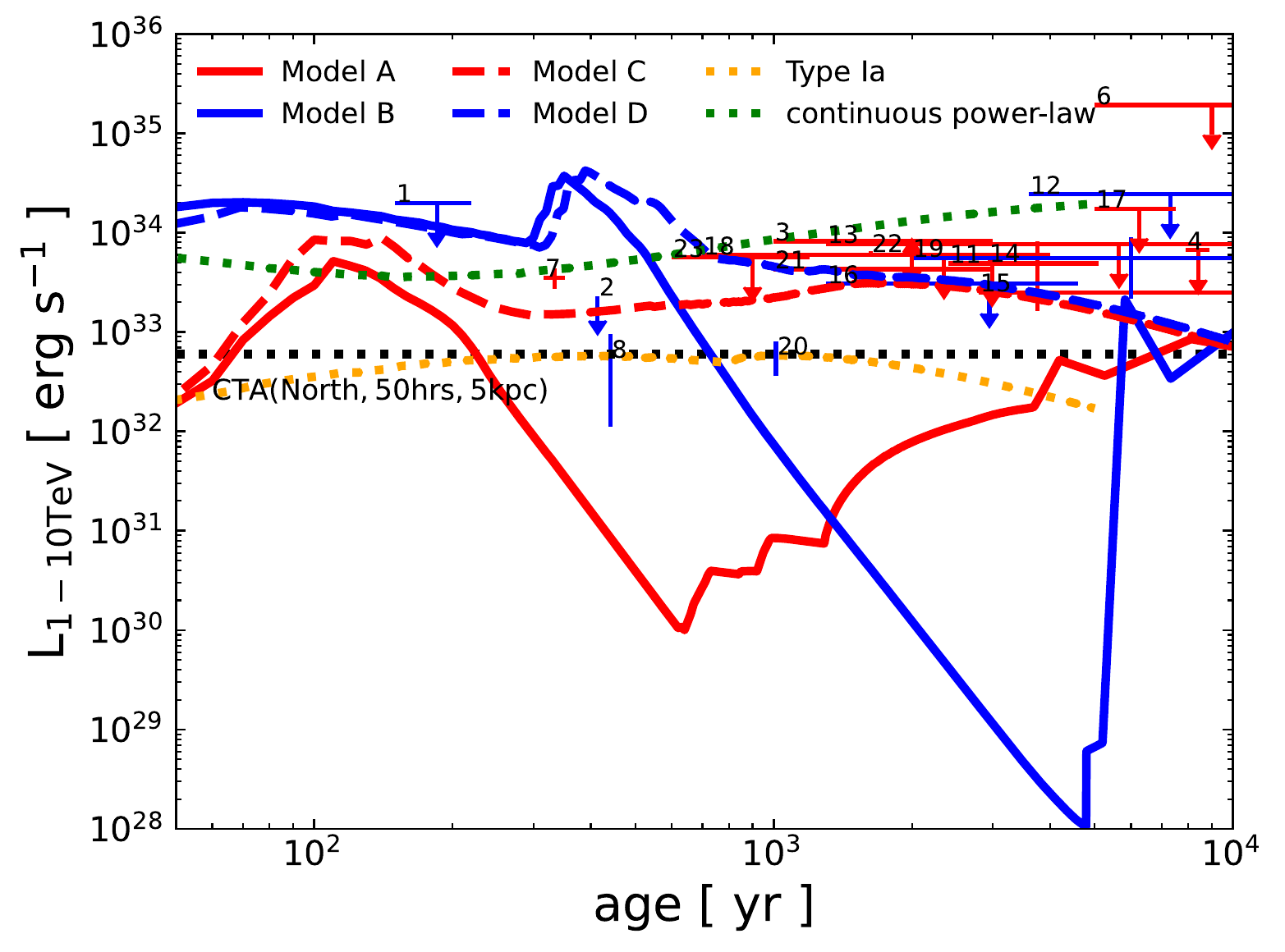}{0.33\textwidth}{(c)}}
\caption{Light curves of the 1~GHz radio continuum (left panel (a)), $\gamma$-ray integrated over the 1-100 GeV band (middle panel (b)) and 1-10 TeV band (right panel(c)). The line formats are the same as in Fig.~\ref{fig:RV_t}. In panel (a), (b) and (c), the detection limit of VLA, {\it Fermi}-LAT and CTA are plotted with black lines, respectively. Results from multi-wavelength observations of selected SNRs as shown in Fig.~\ref{fig:RV_t} are also overlaid. \label{fig:lum}}
\end{figure*}

\begin{figure*}[ht!]
\plotone{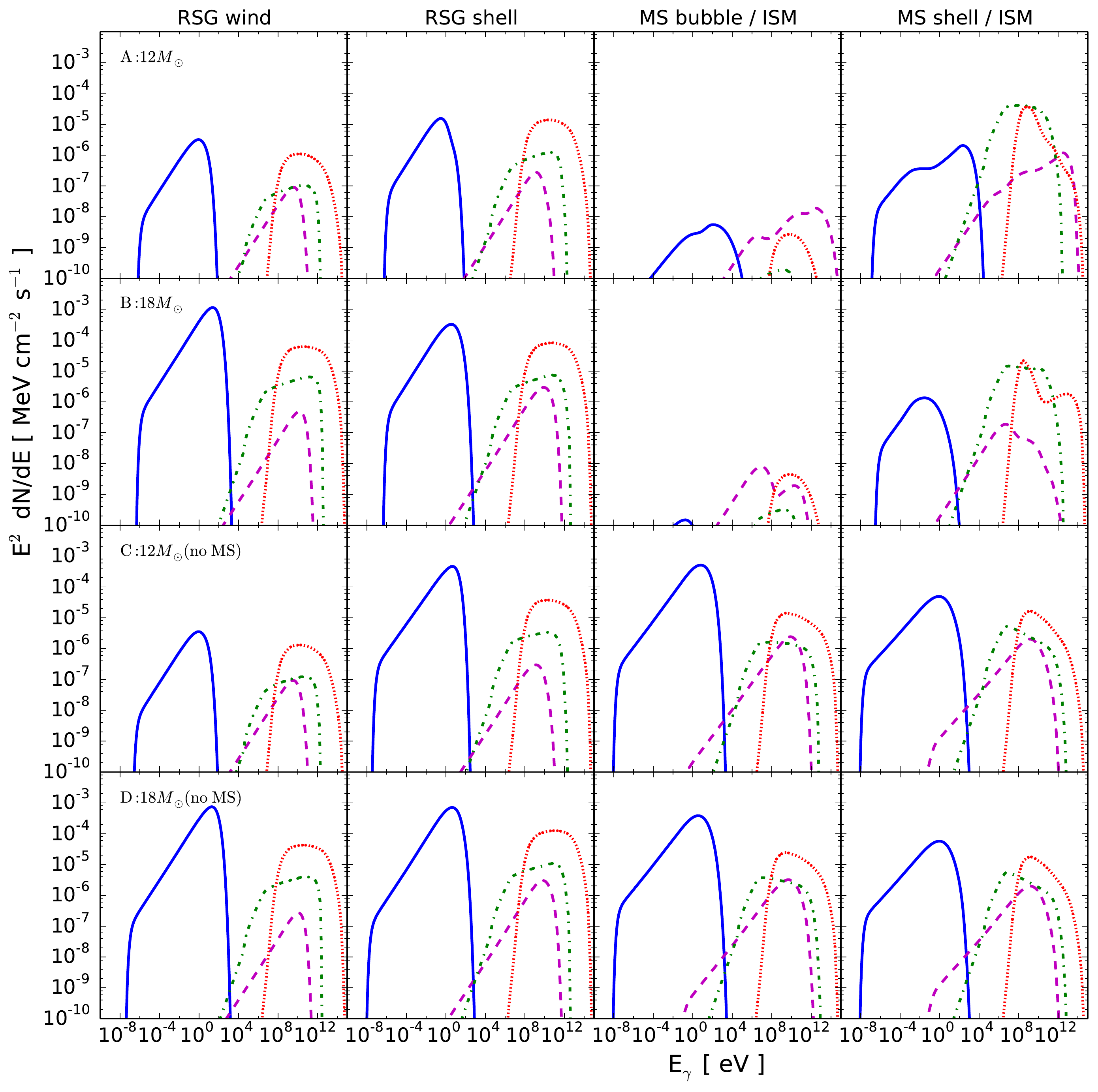}
\caption{Broadband SED from a Type II SNR with different progenitor masses and CSM models (top to bottom) and at different ages (left to right). The exact ages for each of the four panels from left to right are characterized by the location of the forward shock in different regions of the CSM environment, and are showed in panel (b) of Fig.~\ref{fig:lum} with red arrows for model A and C, and blue arrows for model B and D. The emission components include synchrotron (blue solid), $\pi^0$ decay (red dotted), IC (magenta dashed), and non-thermal bremsstrahlung (green dot-dashed).\label{fig:SED}}
\end{figure*}

Figure.~\ref{fig:RV_t} shows the time evolution of the SNR radius $R_\mathrm{sk}$ (upper panel) and shock velocity $V_\mathrm{sk}$ (lower panel) for each of our models. As a reference, we also plot the results of two fiducial models from YL19\footnote{As the SN ejecta, an exponential profile $\rho(r)\propto \exp{(-r/r_\mathrm{ej})}$ for Type Ia case \citep{1998ApJ...497..807D} was assumed. For continuous power-law model, a power-law envelope model was used with the same eqn.~(\ref{eq:ejecta}). The ejecta mass and kinetic energy of each case were $1.4\ M_\odot$ and $10^{51}\ \mathrm{erg}$ for Type Ia case, and $3.0\ M_\odot$ and $10^{51}\ \mathrm{erg}$ for power-law case. $n_\mathrm{SN}=7$ was also assumed in the latter model.} (see their models A2 and B2), i.e., a model with a uniform ISM-like environment with $n_\mathrm{ISM}=0.1\ \mathrm{cm}^{-3}$ (hereafter, ``Type Ia" case) and another with a power-law CSM extended to an infinite radius with $\dot{M}=10^{-5}\ M_\odot/\mathrm{yr}$ (hereafter, ``continuous power-law" case). Observational data from a selection of $\gamma$-ray bright SNRs are also plotted with blue points for Type Ia SNRs and red points for core collapse SNRs. They are sorted with numbers and the corresponding table is summarized in Fig.~11 in YL19. The details and references for the observational data can again be found in YL19. 

We first look at the results from model D (blue dashed line) which has the most straightforward evolution behavior. In the early phase with $t\le300\ \mathrm{yr}$, the SNR shock is propagating inside the RSG wind, and the time evolution is similar to the continuous power-law case except that the absolute values are slightly different because parameter values such as the ejecta mass are not the same. As the SNR continues to expand, it collides with the RSG shell and results in a small deceleration of the shock. The deceleration is not significant because the mass inside the RSG shell is much smaller than the ejecta mass. Finally, the SNR expands into the uniform ISM region and eventually sweeps up an amount of ISM material more massive than the ejecta, and the SNR enters the self-similar Sedov phase. During this phase, the shock radius and velocity depend only on the SN explosion energy, ISM gas density and age, therefore model D shows a similar behavior to the Type Ia case after $t\ge3,000\ \mathrm{yr}$. For model C (red dashed line), the shock decelerates at an earlier time than model D because the RSG shell is located at a smaller radius than in model D for the reasons already explained in section.~\ref{Method:CSM}. Otherwise, the general evolution is qualitatively similar to model D.

Model B follows the same evolution trend as model D until the shock hits the RSG shell. The shock breaks out from the RSG shell into a tenuous MS bubble, so that the shock accelerates and the expansion of the SNR speeds up. Afterwards, the shock collides with a dense cold shell at the outer edge of the MS bubble, and rapidly decelerates to $V_\mathrm{sk}\sim10\ \mathrm{km/s}$. The expansion of the SNR then slows down drastically and the SNR size stays more-or-less unchanged. The evolution shown by model A is qualitative similar to model B except for differences in timing simply due to the different locations of the MS bubble.  

Figure.~\ref{fig:lum} shows the light curves for the 1 GHz radio continuum (left panel (a)), GeV $\gamma$-rays in the 1-100 GeV band (middle panel (b)), and TeV $\gamma$-rays in the 1-10 TeV band (right panel (c)). The color and line formats are the same as Fig.~\ref{fig:RV_t}. From left to right, Figure.~\ref{fig:SED} shows the spectral energy distribution (SED) from each model at four chosen characteristic ages as indicated by the arrows in panel (b) of Fig.~\ref{fig:lum} (red arrows for model A and C with $M_{\rm ZAMS} = 12M_\odot$, and blue arrows for models B and D with $M_{\rm ZAMS} =18M_\odot$). 

The light curves from model D behave similarly in all wavelengths to the continuous power-law case at early times ($t\le300\ \mathrm{yr}$) and to the Type Ia case at larger ages ($t\ge3000\ \mathrm{yr}$) which is in accordance with the hydrodynamical evolution. At early times, the $\gamma$-rays are dominated by the hadronic component from $\pi^0$ decay because of the high gas density in the RSG wind, and suffer from strong adiabatic loss due to the inverse power-law distribution of the CSM as $r^{-2}$. As a result, the $\gamma$-ray luminosity decrease with time. The shock expands into the uniform ISM later on, and the $\gamma$-rays stay dominated by the $\pi^0$ decay channel. The spectral power-law index of the accelerated proton and hence the $\gamma$-ray spectrum becomes steeper however due to shock deceleration in the ISM and an increased influence from the Alfv\'en velocity on the non-linear DSA process as the SNR enters its Sedov phase (see the rightmost panel in Fig.~\ref{fig:SED}), and the $\gamma$-ray luminosity decreases accordingly in particular for the TeV band. These evolution behaviors are found to be similar to the results in YL19. At intermediate ages ($300\le t\le3000\ \mathrm{yr}$), the SNR hits the RSG shell, and the emissions brighten briefly for about 200 yrs before the light curves gradually converge back to those similar to the Type Ia case. In model C, the SNR collides with the RSG shell at an earlier age of 60 yrs and brightens from 100 to 200 yrs, but otherwise shows similar behavior to model D after an age of 2000 yrs. 

Of the biggest interest and surprise are the results from model B. Up until the collision of the SNR with the RSG shell ($t\le500\ \mathrm{yr}$), the light curves basically follow the same evolution as model D. After the collision, however, the radio and $\gamma$-ray luminosities rapidly decrease to a point that they are undetectable by current observational instruments. We can interpret this rapid dimming based on two reasons. First, as the SNR shock enters the tenuous and hot MS bubble region, it becomes difficult for the shock to accelerate particles through DSA because injection becomes inefficient due to the low density of the ambient gas $n\sim10^{-4}\ \mathrm{cm}^{-3}$, and the shock sonic Mach number $M_s$ decreases drastically due to the high temperature $T\sim10^8\ \mathrm{K}$ in the bubble, namely $M_s=V_\mathrm{sk}/C_s\sim5\ (V_\mathrm{sk}/5\times10^3\ \mathrm{km/s})(T/10^8\ \mathrm{K})^{-1/2}$, where $C_s$ is the local sound speed. Second, the SNR expands rapidly while the shock is inside the MS bubble. The particles accelerated earlier on in the RSG wind suffer from fast adiabatic loss from the rapid expansion, and the luminosities drop down by at least three orders of magnitudes. These results can also be observed from the SEDs in the third column in Fig.~\ref{fig:SED}. After the SNR shock has propagated through the bubble and eventually hit the cold dense shell at the edge, the shock start to sweep up the dense material in the shell and the non-thermal emissions are then enhanced from the increased gas density. The SNR brightens again enough to be observable by currently available detectors, as will be discussed in more details below. 

The SNR shock is interacting with the MS shell at an age of 10,000 yr (Fig.~\ref{fig:RV_t}). After that, it is expected that the shock will break out from the shell and propagate into the uniform ISM region. In this phase, the shock velocity should have decelerated to a velocity too low to accelerate new particles efficiently in the ISM, and the luminosities will decrease with time due to adiabatic loss. Continuing our simulations beyond 10,000 yrs would allow us to estimate the exact lifespan of the SNR in the radio and $\gamma$-ray energy bands, but it is beyond the scope of this work.  

Model A shows slightly different results from model B, in particular during the MS bubble phase. The ejecta mass of model A is smaller than model B, and the total mass inside the RSG wind is also about 5 times smaller. This leads to a shock velocity in model A almost 2 times higher than in model B when the shock is inside the MS bubble (Fig.~\ref{fig:RV_t}). As a result, the sonic Mach number is also higher by roughly a factor $\sim$ 2 at $M_s\sim10$ while inside the MS bubble. This shock can accelerate new particles despite the low gas density inside the bubble, therefore the light curves rise gradually with time from 600 yrs which is different from the behavior shown by model B with a more massive progenitor. 

To assess the observational detectability of a Type II SNR based on our models, observation sensitivities in the radio and $\gamma$-ray bands are plotted in panel (a), (b) and (c) in Fig.~\ref{fig:lum} with black dotted lines. We compare the detection limit of the Very Large Array (VLA) with our models for the radio band. Radio galaxies and active galactic nuclei are often observed with a sensitivity $\sim100\ \mu\mathrm{Jy}$ at 1.4 GHz \citep[e.g. ][]{2004AJ....128.1974S,2012MNRAS.421.3060S}. The lower limit of the radio luminosity from a source at a distance of 10 kpc is therefore $\sim2\times10^{28}\ \mathrm{erg/s}$. We note that it is a very optimistic limit, since this is the typical sensitivity for a targeted observation. If there is no detection in other wavelength, the radio sensitivity should be lower. We also compare with the sensitivity of the {\it Fermi} Large Area Telescope ({\it Fermi}-LAT) for GeV $\gamma$-rays. For TeV $\gamma$-rays, we use the sensitivity data of the Cherenkov Telescope Array (CTA), the most powerful next-generation ground-based $\gamma$-ray telescope expected to start observing the Universe in year 2022 \citep{2019scta.book.....C}. {\it Fermi}-LAT has a flux sensitivity of $\sim 2\times10^{-12}\ \mathrm{erg\ cm^{-2}\ s^{-1}}$ in the 1-100 GeV band based on 10 yrs of survey data (see, for details, \cite{2020ApJS..247...33A,2020arXiv200511208B} and \url{https://www.slac.stanford.edu/exp/glast/groups/canda/lat_Performance.htm}), which corresponds to a luminosity $\sim1.2\times10^{34}\ \mathrm{erg/s}$ for a $\gamma$-ray source at 5 kpc. The detection limit of CTA at 5 kpc is $\sim6\times10^{32}\ \mathrm{erg/s}$ with a flux sensitivity $\sim 10^{-13}\ \mathrm{erg\ cm^{-2}\ s^{-1}}$ in the 1-10 TeV band for the northern telescopes and an observation time of 50 hrs (see, for details, \url{http://www.cta-observatory.org/science/cta-performance/ (version prod3b-v1)}). We do not consider other effects like interstellar absorption and source contamination for simplicity.

Our results show that the $\gamma$-rays cannot not be observed from 1000 yr to $10^4$ yrs for the case with a $18M\odot$ progenitor, and from 300 yr to $10^4$ yr for a $12M\odot$ star. In addition, the radio emission also stays faint and barely comparable to the VLA sensitivity limit until 5000 yrs. On the contrary, we can observe Type II SNRs with ages of $5000\ \mathrm{yr}\le t\le10000\ \mathrm{yr}$ but only in the radio. So, we conclude that with the presence of a tenuous hot bubble created by the MS stellar wind, most Type II SNRs experience a ``dark age'' in which they become too faint to be observable at ages $\sim 1000 - 5000$ yrs, although the span and exact timing can depend on the surrounding environment, mass loss history of individual progenitors and the detection limits of currently available detectors.

\section{Discussion} \label{sec:Discussion}
We have chosen a few model parameters related to DSA to match our previous model of RX J1713 (see, e.g., Fig.3 in YL19), which showed a good agreement with the bulk properties and the overall broadband spectrum but without considering a collision with molecular clouds. However, the correlation of RX J1713 with molecular clouds has been reported by some recent works \citep[e.g.,][]{2012ApJ...746...82F,2020ApJ...900L...5T}, which may necessitate a revision of our model for this particular object in the future. Our results and conclusions in this work are mainly dependent of the bulk dynamics of the SNR shock in its surrounding CSM environment created by the RSG progenitors, which do not rely on any fine-tuning of model parameters mentioned above. Therefore, our results can be considered robust and present two possibilities:
\begin{enumerate}
    \item If the MS bubbles exist, most Type II SNRs cannot be detected as it enters the bubble, which corresponds to an age of $10^3$ - $5\times10^3\ \mathrm{yr}$ for a RSG progenitor exploded inside a typical ISM,
    \item The MS bubbles indeed might not exist or be compact enough so that accelerated particles are not affected too much by adiabatic loss. 
\end{enumerate}
If the first scenario is true, all detected core collapse SNRs so far with ages around 1000 to 5000 yrs old are most probably not originated from Type II SNe. Indeed, the total SN rate in our Galaxy is almost $1/30\ \mathrm{yr}^{-1}$ \citep[e.g,][]{2013ApJ...778..164A} so that the number of expected SNRs with an age of 1000 to 5000 yrs should be at least 100. Nevertheless the number of SNRs detected in radio and other wavelengths falling into this age range is only at an order of ten \citep{2017yCat.7278....0G,2016ApJS..224....8A,2018A&A...612A...3H}. Because Type II SNe are expected to produce almost half of the total population of SNRs \citep[e.g.][]{2007ApJ...661.1013L}, this is consistent with our results that many Type II SNRs actually cannot be detected. On the contrary, our results for cases without the MS bubble show that the SNRs are bright enough to be detected with present detectors. The detection rate should be larger if the MS bubbles do not exist or compact enough to be unimportant. The interpretation therefore depends on the general (non-)existence of MS bubbles around the massive star progenitors.

 One related caveat is that we have only considered a simple scenario for stellar evolution in this work. For example, the wind velocity plays an important role for shaping the CSM environment. If the MS stellar wind is slower than what we assumed here, and/or the RSG wind is faster, the MS bubble is expected to be smaller in size so that the RSG wind can sweep through almost its entirety before core collapse. A smaller mass loss in the MS phase will lead to the same result. From this point of view, type Ib/c SNe are possibly important objects. The progenitors of type Ib/c SNe are thought to be Wolf-Rayat (WR) stars. A WR star is a compact star which has lost its entire hydrogen envelope via stellar wind and/or binary interaction through a phase of Roche-lobe overflow. It ejects very fast wind with $V_\mathrm{w}\sim10^3\ \mathrm{km/s}$, and this wind can sweep up the MS bubble all the way close to the edge where the dense cold shell sits. This may help their SNRs avoid the strong adiabatic loss of the accelerated particle due to a fast expansion of the remnant in the MS bubble. This therefore may present a possibility that most of the detected core collapse SNRs with an age of a few 1000 yrs are coming from stripped envelope SNe. We are now expanding our study to calculate models for SNRs from a type Ib/c origin to explore this possibility. The results will be reported in a separate paper in the near future. 

Another caveat is that it is possible that some of the progenitors are evolving inside or close to an environment with a higher density than the average ISM, for example, giant molecular clouds (MCs). In these environments, the MS wind can sweep up a large amount of gas in the surrounding dense gas and rapidly converts its kinetic energy to thermal energy, halting its expansion effectively \citep{2015A&A...573A..10M}. In addition, the emission luminosity is also expected to be higher because of the high density. 
However, these SNRs are exploded in a small cavity surrounded by a dense environment, so they are expected to enter the radiative phase quickly and become very dim (so-called ``dark SNRs"), and their lifespans will be relatively short anyway.

Anyhow, the detection of MS bubbles around SN progenitors is indispensable for a resolution. However, that is quite difficult because MS bubbles typically have very low densities and high temperatures, so that both emission and absorption are inefficient. \cite{2017MNRAS.466.1857G} reported a first example of MS bubble detection. By a comparison to radiation-hydrodynamics simulations, they interpreted the observation by the collision of the MS wind from B type stars and nearby MCs. While illuminating, a statistical discussion of MS bubble is still impossible due to the small sample of observational examples. Theoretical approaches is therefore important. An expansion of our work to consider higher density environments will be done in a follow-up paper.  

At last, we note that our simulations are 1-dimensional and do not include multi-dimensional effects. This imposes that the ISM is isotropically distributed. If the wind material and ISM distribute anisotropically, and/or the SN exploded asymmetrically, a non-spherical situation is expected, probably accompanied by bow shocks \citep[e.g.,][]{2012A&A...541A...1M}. Multi-dimensional effects like Rayleigh-Taylor fingers have been observed too in a number of remnants like Tycho \citep[e.g.,][]{2005ApJ...634..376W} which can also affect the emission to some extent. To investigate these effects, especially for the modeling of specific objects, multi-D simulations will indeed be desirable. As a first study, however, we aim at constructing a “standard” evolutionary picture for Type II SNRs in general, and evaluate the effects of the (non-)existence of a rarefied MS bubble beyond the RSG wind on the bulk properties of the non-thermal emission. In this context, we consider a parametric study using 1-D simulations suitable.

\section{Conclusion} \label{sec:Conclusion}
Young SNRs are usually bright in multi-wavelength from radio to $\gamma$-ray from the interaction between CRs accelerated by the SNR shock and the surrounding ambient environment. This suggests that non-thermal emissions from SNRs are effective probes of the CSM structure and hence the mass-loss history of SN progenitors. In this work, we have calculated the long-term time evolution of non-thermal emissions from Type II SNRs interacting with a realistic CSM considering stellar evolution and mass-loss history of their progenitors. 

We show that the non-thermal emissions are bright enough to be observed by current and future detectors in the RSG wind phase ($t\le1000\ \mathrm{yr}$), but become very faint beyond detectable in the MS bubble phase ($1000\ \mathrm{yr}\le t\le5000\ \mathrm{yr}$). After the collision with the MS shell ($t\ge5000\mathrm{yr}$), the SNR re-brightens in radio and $\gamma$-rays, but gradually declines in luminosity immediately afterwards due to a rapid deceleration of the shock in the dense cold shell. We conclude that most Type II SNRs experience a ``dark age" from 1000 to 5000 yrs for progenitors with ZAMS mass $M_\mathrm{ZAMS}\le18\ M_\odot$ exploded in a typical ISM surrounding. This phenomenon is mainly caused by an inefficient particle acceleration and fast adiabatic loss in the thin and hot MS bubble. Our results may help to fill in the gap between the Galactic SN rate and SNR observations. While the existence of a spatially extended MS bubble around massive stars is still uncertain, and is affected by various factors such as the wind properties, the surrounding ISM environment and so on, our conclusion is robust in that it does not depend on any fine-tuning of parameters of aspects such as particle acceleration and explosion properties. A further investigation by expanding our parameter space including different progenitor systems is under way and will be reported in a follow-up work.

\begin{acknowledgments}
H.Y. acknowledges support by JSPS Fellows Grant No.JP20J10300. S.H.L. acknowledges support by JSPS Grant No. JP19K03913 and the World Premier International Research Center Initiative (WPI), MEXT, Japan. K.M. acknowledges support from JSPS KAKENHI grant JP18H05223, JP20H04737 and JP20H00174.
\end{acknowledgments}


\bibliography{reference}{}
\bibliographystyle{aasjournal}



\end{document}